\documentclass[12pt,letter]{article}
\usepackage{graphicx}
\usepackage{amsmath}
\usepackage{amssymb}
\usepackage{cancel}
\usepackage[left=2cm,right=2cm,top=3cm,bottom=2cm]{geometry}
\usepackage{setspace}
\usepackage{float}
\usepackage[super,sort&compress,comma]{natbib}
\usepackage[font={small}]{caption}
\usepackage[labelfont=bf]{caption}
\usepackage{bm}
\usepackage[T1]{fontenc}
\usepackage{microtype}
\usepackage{subfigure}
\usepackage{chemformula}
\usepackage[version=4]{mhchem}
\usepackage{textcomp}

\usepackage{color}
\include{MyCommand}

\usepackage{tabularx}
\usepackage{booktabs}
\usepackage{multirow}
\usepackage{mathtools}
\usepackage{bm}
\usepackage{gensymb}
\usepackage{esvect}
\usepackage{plimsoll}

\title{The Role of Interfaces and Charge for Chemical Reactivity in Microdroplets}

\begin{document}
\date{}
\maketitle
\begin{center}
\author{R. Allen LaCour$^{1,2}$*, Joseph P. Heindel$^{1,2}$*,  Ruoqi Zhao$^{1,2}$, Teresa Head-Gordon$^{1,2,3}$}

$^1$Kenneth S. Pitzer Theory Center and Department of Chemistry\\
$^2$Chemical Sciences Division, Lawrence Berkeley National Laboratory\\
$^3$Departments of Bioengineering and Chemical and Biomolecular Engineering\\
University of California, Berkeley, CA, USA

corresponding author: thg@berkeley.edu

*Equal first authors
\end{center}
\begin{abstract}
\noindent
A wide variety of reactions are reported to be dramatically accelerated in aqueous microdroplets, making them a promising platform for environmentally clean chemical synthesis. However to fully utilize the microdroplets for accelerating chemical reactions requires a fundamental understanding of how microdroplet chemistry differs from that of a homogeneous phase. Here we provide our perspective on recent progress to this end both experimentally and theoretically. We begin by reviewing the many ways in which microdroplets can be prepared, creating water/hydrophobic interfaces which have been frequently implicated in microdroplet reactivity due to preferential surface adsorption of solutes, persistent electric fields, and their acidity or basicity. These features of the interface interplay with specific mechanisms proposed for microdroplet reactivity, including partial solvation and possible gas phase channels.
We especially highlight the role of droplet charge, which appears key to understanding how certain reactions, like the formation of hydrogen peroxide and reduced transition metal complexes, are thermodynamically possible in microdroplets. 
Lastly, we emphasize opportunities for theoretical advances in the microdroplet field generally, and to suggest experiments which would greatly enhance our understanding of this fascinating and emerging subject.
\end{abstract}

\newpage
\section{Introduction}
Many organic and redox reactions are reported to occur with much faster kinetics in water microdroplets and oil-water emulsions than in bulk solution. 
Although ``on-water'' reactions\cite{narayan2005water, chanda2009organic} and reactivity in atmospheric aerosols\cite{graedel1981chemistry,Prather2008} have been studied longer and are well-established, the first reports on reactivity in laboratory-prepared aqueous microdroplets appeared only within the past two decades\cite{enami2008acidity, girod2011accelerated, perry2011detecting}.
In early studies, which mainly concerned  microdroplets prepared through electrospray ionization (ESI)\cite{fenn1989electrospray, banerjee2012electrospray}, many types of organic reactions were found to be accelerated\cite{girod2011accelerated, badu2012accelerated, perry2011detecting, muller2012accelerated, banerjee2015syntheses,  li2014synthesis, lee2015microdroplet, nam2017abiotic, banerjee2017can, yan2021emerging, lee2015acceleration, wei2020accelerated}, indicating that microdroplets may be generally useful vessels for organic synthesis. Subsequently, acceleration was also reported in related interfacial  environments, including water-in-oil emulsions\cite{PhysRevLett.112.028301, vogel2020corona}, thin films\cite{badu2012accelerated, yan2013chemical, wei2017reaction}, and levitated droplets\cite{li2018accelerated, li2021reaction, abdelaziz2013green, crawford2016real, bain2016accelerated, jacobs2017exploring}. Especially interesting is the wide variety of redox chemistry reported to occur in aqueous microdroplets\cite{lee2018spontaneous, lee2019micrometer, gong2022spontaneous, qiu2022simultaneous, song2022spraying, chen2023spontaneous, yuan2023spontaneous, xing2022spontaneous, zhang2022high, jin2023spontaneous}, including the reduction of various metals\cite{yuan2023spontaneous, he2022vapor} and the oxidation of water into hydrogen peroxide\cite{lee2019micrometer, lee2020condensing, musskopf2021air,mehrgardi2022sprayed}.
Intriguingly, many of these reactions are thermodynamically unfavorable in bulk water.

However, explaining the origins of microdroplet reactivity has proven challenging, with the inherent complexity of the multiphase system leading to many plausible mechanisms for rate acceleration\cite{yan2016organic, qiu2021reaction, chamberlayne2020simple,  xiong2020strong1, hao2022can, heindel2022spontaneous, xiong2020strong2, ruiz2022disentangling, jacobs2018studying, rovelli2020critical, chen2023role}.
Due to the high surface-to-volume ratio of the microdroplets, most might agree that the interface likely plays a key role. Problematically, however, many aspects of water interfaces that are relevant for reactivity, such as their acidity or basicity\cite{zimmermann2010hydroxide}, have themselves proven challenging to understand\cite{agmon2016protons}.
Another important, but perhaps less appreciated, feature of microdroplets is their charge\cite{banerjee2015syntheses, chamberlayne2020simple, rovelli2020critical, heindel2024role}, which is extremely relevant for reaction kinetics and thermodynamics\cite{colussi2023mechanism, heindel2024role}. 
This is an important clue as to what type of microdroplet preparation techniques generate charged or uncharged droplets,  and in which of these cases is there observed accelerated reactivity.

Here we provide our perspective and understanding of the underlying reasons for chemical reactivity in microdroplets and related systems like oil-water emulsions. We begin by reviewing the experimental landscape for creating droplets, measuring their reactivity, and provide a brief catalogue of reactions that have been reported to be accelerated in droplets relative to the bulk phase. Next we discuss the current understanding of water/hydrophobic interfaces, highlighting areas of contention that may be relevant for microdroplet reactivity, like the recent attempts to spectroscopically characterize emulsion interfaces\cite{pullanchery2021charge,shi2024water} or test spontaneous formation of hydrogen peroxide\cite{mehrgardi2022sprayed,eatoo2024busting}. We also discuss the selectivity of organic molecules and ions for interfaces, and stress the importance of the strength of interfacial adsorption that dictates the thermodynamics of reactions in uncharged droplets. 

We believe it is of primary importance to consider the role of droplet charge, which substantially alters the reaction thermodynamics of microdroplets.\cite{chamberlayne2020simple,colussi2023mechanism,heindel2024role} Charge is key for redox reactions in particular, since the presence of both positively and negatively charged droplets likely explains the simultaneous reduction and oxidation potential of microdroplets.\cite{qiu2022simultaneous} Charged microdroplets are often created as a result of contact electrification which distinguishes them from bulk liquids where ionic strength, not excess charge, is the appropriate reaction variable. Finally, we critically assess contributions from other proposed mechanisms, while also highlighting their connections. For example, the presence of electric fields at hydrophobic-water interfaces can be regarded as a unified measure of molecular interactions which provides information about interfacial structure and whether or not a droplet is charged. We conclude our perspective by discussing remaining open questions about microdroplet reactivity, with suggestions for future experimental and theoretical studies that would advance this fascinating subject.

\section{Preparing Microdroplets and their Reactivity}
\begin{figure}[H]
    \centering
    \includegraphics[width=0.99\textwidth]{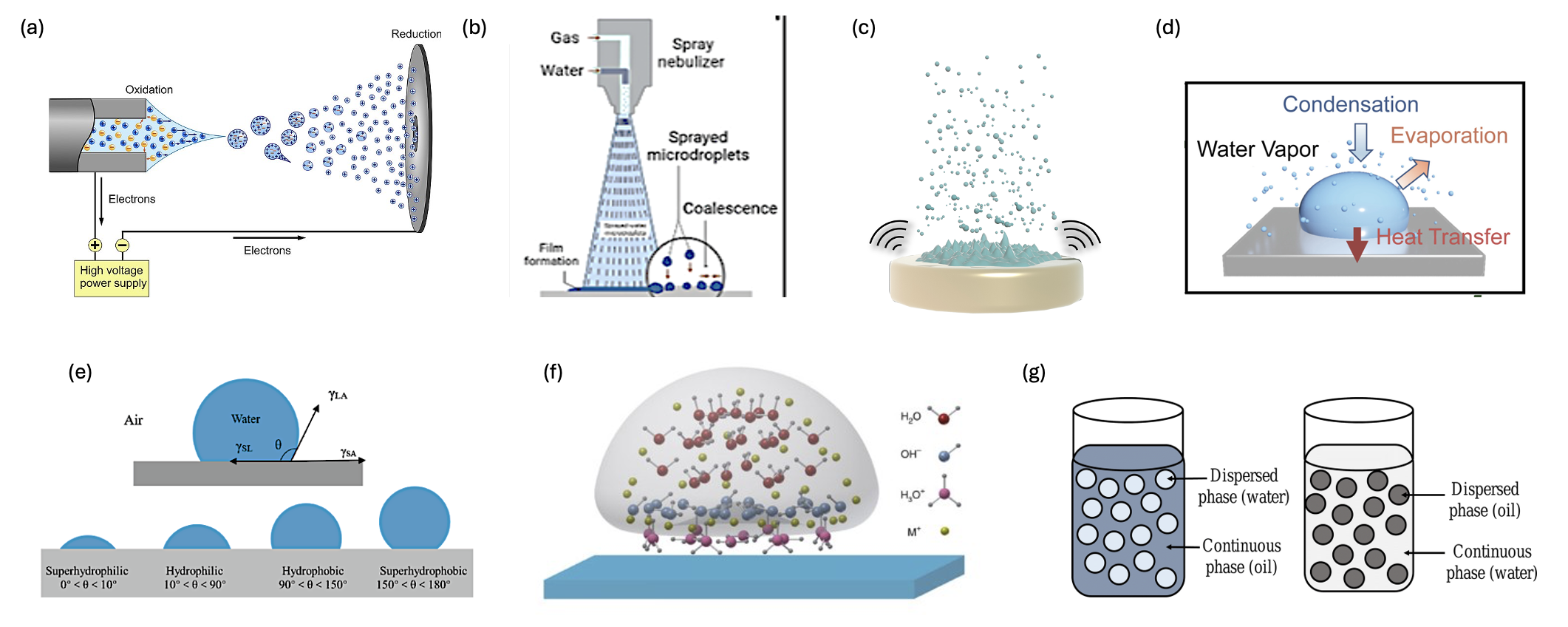}
    \caption{\textit{Different ways of producing microdroplets.} (a) electrospray ionization.  Adapted from wikipedia (b) gas nebulization. Adapted from reference \cite{Mehrgardi2024} (c) ultrasound humidification. Adapted from reference \cite{lin2023size} (d) water condensation. Adapted from reference \cite{lee2020pnas} (e) deposited droplets. Adapted with permission from ref. \cite{Liu2021} (f) Levitated or Leidenfrost droplet. Adapted from reference \cite{abdelaziz2013green} (g) oil-in-water and water-in-oil emulsions. Adapted with permission from ref. \cite{Sawant2021}}
    \label{fig:techniques}
\end{figure}

Figure \ref{fig:techniques} shows some of the different experimental techniques and conditions used for creating microdroplets. To draw out these distinctions more specifically, we discuss several methods for microdroplet formation as shown in Figure \ref{fig:techniques}: (a) electrospray ionization, (b) gas nebulization, (c) ultrasonic humidifcation, (d) water condensation from the vapor, (e) deposited droplets, (f) levitated droplets, and (g) oil-water emulsions. We take each of the microdroplet preparations in turn to describe how the technique leads to generation of droplets, and their observed properties, as it is important to connect their features to observed reactive chemistry.

\textbf{Electrospray Ionization.} Most early experiments used variants of electrospray ionization (ESI), in which charged aerosols are produced by applying high voltage to the liquid. 
These charged microdroplets demonstrate unique chemistry thats differs dramatically from the aqueous phase.
Examples include the acceleration of many common organic reactions\cite{muller2012accelerated, banerjee2015syntheses, nam2017abiotic, nam2018abiotic} , the acceleration of acid- and base-catalyzed reactions in positively and negatively charged microdroplets, respectively\cite{girod2011accelerated}, and the formation of nanoparticles\cite{li2014synthesis}.
Bannerjee et al. \cite{banerjee2012electrospray} found the rate of several reactions to increase with applied voltage, highlighting the role of droplet charge.
Furthermore, the species responsible for the droplet's charge, including hydronium in positively charged droplets and hydroxide in negatively charged ones,\cite{banerjee2012electrospray} will be present in excess without counterions (as in the bulk water phase), influencing chemistry.
These issues make the droplet surface created by electrospray unrepresentative of generic air-water interfaces\cite{wilson2020kinetic} and may drive reactions via a pathway independent of
the interface.

\textbf{Gas Nebulization.} Gas nebulization, in which gas flows into a stream of water to generate a fine mist of droplets, has been used extensively in microdroplet chemistry.
The size of the droplets can be loosely controlled by varying the flow rate of the nebulizing gases (Figure \ref{fig:techniques}). Gao et al. used gas nebulization to show that the Dakin and Baeyer-Villiger reactions proceed without the addition of peroxides, which are necessary catalysts in the bulk aqueous phase.\cite{gao2019} Spontaneous reduction of several organic molecules was found to occur in microdroplets prepared by gas nebulization\cite{lee2019jacs}: pyruvate to lactate, lipoic acid to dihydrolipoic acid, fumarate to succinate, oxaloacetate to malate, and the formation of both pyridyl anions and hydroxypyridine in microdroplets of a water/pyridine mixture.\cite{Zhao2022} 
From these observations it is speculated that it is the availability of free electrons and oxidative species such as \ce{OH^.} that can simultaneously reduce and oxidize to create products or to provide the needed hydrogen peroxide catalyst.\cite{Zhao2022,lee2019jacs} Gas nebulization also creates charged droplets through contact electrification\cite{wang2019origin,chen2022water}, a topic which we discuss in more detail below. 

\textbf{Ultrasonic Humidification.}  In ultrasonic humidification, a mist is created by mechanically vibrating a liquid in the kHz to MHz frequency ranges. The resulting mist contains droplets as small as $\sim$1 $\mu$m in diameter. 
The droplets appear to be charged, with larger and smaller droplets more likely to be positively or negatively charged, respectively\cite{lin2023size}. Using ultrasonic humidification, the Mishra lab\cite{musskopf2021air} found spontaneous formation of $\sim$ 1 $\mu$M H$_2$O$_2$.

Nguyen and Nguyen argue that ultrasonication results in cavities in the fluid that collapse, releasing energy sufficient to produce reactive species such as \ce{OH^.}, \ce{H^.}, \ce{HO_2^.}, which eventually combine to form \ce{H2O2}.\cite{nguyen2022revisiting} 
Indeed, many highly unfavorable reactions have been observed to occur during cavitation.\cite{suslick2008inside} 
In support of this argument, Nguyen demonstrated that doping the microdroplets with ions known to prefer the interior of a microdroplet, such as \ce{SCN^-}, quenches \ce{H2O2} formation, as does the addition of \ce{HCl}. Thus, sonication may also drive reactions via a pathway independent of the air/water interface. Recently, it has been claimed that only water containing dissolved O$_2$ using gas nebulization, and subsequent collection of the product at the solid–water interface, forms H$_2$O$_2$.\cite{eatoo2024busting} These conflicting experimental results indicate the sensitivity of the reactive outcome to details of droplet preparation and collection methods.

\textbf{Water Condensation.} Gently heating water at 50-70 $^\circ$C and condensing the vapor onto a cold surface is likely the most benign way to create microdroplets. Although Lee et al.\cite{lee2020condensing} originally reported that water condensation yielded fairly high concentrations of \ce{H2O2} formation, more recent work\cite{gallo2022formation, musskopf2021air} showed no detectable \ce{H2O2}. 
The lack of measurable reactivity relative to the other droplet methods may be because this method produces droplets larger than $\approx\mathrm{1~ \mu m}$, which are known to be less reactive, and because this preparation does not yield charged droplets. Under these conditions, Eatoo and Mishra\cite{eatoo2024busting} also showed no \ce{H2O2} formation was detected with NMR. 

\textbf{Deposited Droplets.}
Small water droplets can also be prepared by depositing a small amount of solution onto a surface. Evaporation concentrates the reactants, so simply letting the droplet evaporate is one strategy for accelerating multi-reagent reactions.\cite{badu2012accelerated}
As an example, Wei et al.\cite{wei2017reaction} found a Claisen–Schmidt synthesis to be accelerated by two orders of magnitude and with much greater yield.
Later, they found similar results for reactions involving a sugar and an amine in a thin film.\cite{wei2018high}.
Li et al.\cite{li2023size} investigated the condensation chemistry of pyruvic acid in deposited droplets in a humidity- and temperature-controlled environment. They found that the reaction rate was proportional to the surface-to-volume ratio of the droplets, indicating that the reaction occurs at the air-water interface.

\textbf{Levitated Droplets.}
Water droplets can be levitated using electric fields, acoustic waves, or the Leidenfrost effect. Doing so provides a convenient platform to study single droplets and their reactivity\cite{jacobs2017exploring, jacobs2018studying}. Like deposited droplets, they are also prone to evaporation if humidity is not controlled. Using acoustic levitation, Crawford et al.\cite{crawford2016real} found the accelerated degradation of pharmaceuticals.
Bain et al.\cite{bain2016accelerated} reported the acceleration of many organic synthesis reactions in Leidenfrost droplets.
Later, Li et al. examined the Krazynski reaction in  Leidenfrost droplets in great detail, finding that surface-active reagents experienced a greater rate acceleration.
Levitated droplets can also acquire a charge, as Abdelaziz et al.\cite{abdelaziz2013green} found for Leidenfrost droplets. Very recently, levitated droplets of millimeter scale have been manipulated to emit a spray of microdroplets which drive the same reactions found to occur in gas nebulization experiments.\cite{li2024atomization} This calls into question the recently claimed necessity of solid-liquid interfaces in microdroplet experiments\cite{eatoo2024busting} and bolsters the view that charged droplets provide alternative reaction pathways than in the bulk.

\textbf{Oil-water emulsions}. All of the above droplet generation techniques create an air-water interface, whereas oil-water emulsions replace the air phase with an oil phase.
Oil-water emulsions can be stable for months, providing a key advantage over other microdroplets platforms.
They have been shown to accelerate interface-active reactions like imine synthesis.\cite{PhysRevLett.112.028301} 
It is well known from electrophoresis experiments that, under an applied electric field, oil droplets move due to the fact that they carry charge, although the origin of this charge is not fully understood. We consider the evidence on whether air-water and oil-water emulsions are to be generically referred to as ``hydrophobic-water'' interfaces that can be used interchangeably. To this end, while we emphasize their common features, we also note that their dissimilarities can give rise to important differences in interfacial chemical reactivity. 
 
\section{The Role of the Hydrophobic-Water Interface in Reactivity }

\subsection{The Structure of Planar Hydrophobic-Water Interfaces}
\begin{figure}[H]
    \centering
    \includegraphics[width=0.99\textwidth]{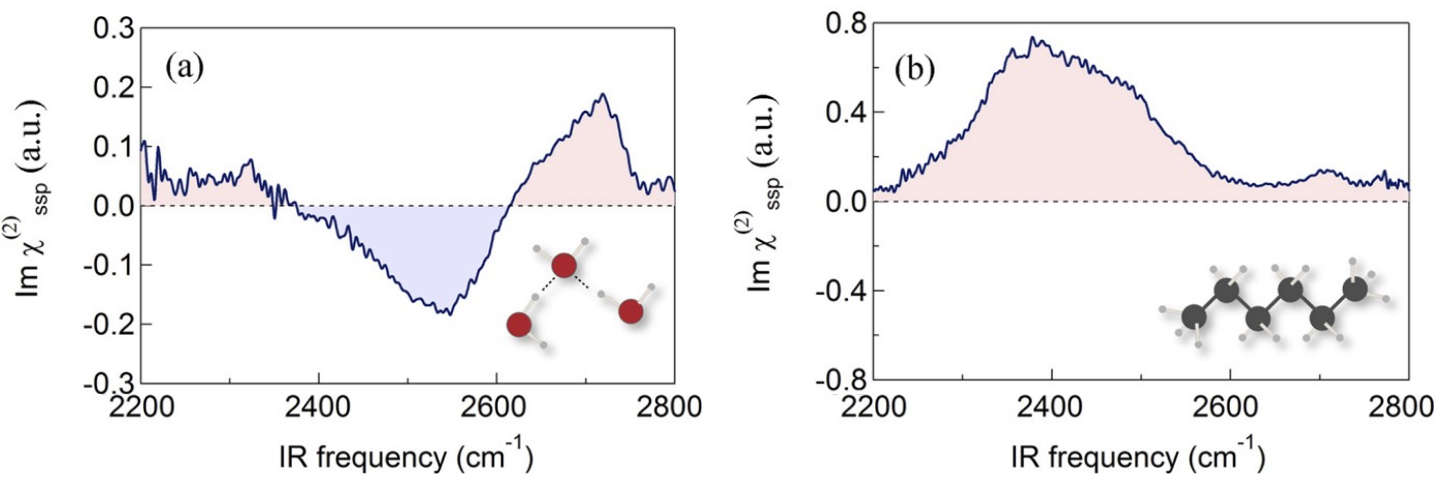}
    \caption{\textit{The heterodyne-collected SFG spectra of two planar hydrophobic interfaces.} (a) air-water interface and b) an oil-water interface. Adapted from reference \cite{strazdaite2015water} with the permission of AIP Publishing.}
    \label{fig:HDSFG}
\end{figure}

A unifying characteristic of water microdroplets is their large interface. 
Consequently, many proposed mechanisms for microdroplet reactivity appeal to the properties of the interface.
Due to the nanoscopic length scales involved, characterizing the interfaces requires advanced experimental techniques\cite{Devlin2024}.
Sum frequency generation spectroscopy (SFG) \cite{du1993vibrational, du1994surface} has proven extremely useful in this regard due to its surface-selective nature\cite{miranda1999liquid, perry2006theoretical, morita2008recent, ishiyama2014theoretical, tian2014recent, tang2020molecular}. Phase-sensitive SFG\cite{ji2008characterization}, which includes heterodyne-detected SFG\cite{nihonyanagi2009direct}, offers the key advantage of the sign indicating whether a vibration's transition dipole points towards the bulk water phase (negative values) or the other phase (positive values). In Figure \ref{fig:HDSFG} we show heterodyne-detected SFG spectra collected by Strazdaite et al.\cite{strazdaite2015water} under the \textit{ssp}-polarization at (a) D$_2$O/air and (b) D$_2$O/hexane planar interfaces.

Inspection of the D$_2$O/air spectra shown in Figure \ref{fig:HDSFG}a yields several insights. 
One is the presence of a positive peak around 2700 cm$^{-1}$ (3700 cm$^{-1}$ in H$_2$O), which has been attributed to the presence of nonbonded DH-(OH-) stretches pointing towards the vapor phase\cite{du1994surface}. Because they are not participating in hydrogen bonds, they are commonly denoted as ``free'' or ``dangling'' stretches.
Second is the presence of a broad negative feature around 2500 cm$^{-1}$ (3400 cm$^{-1}$ in H$_2$O). 
This feature has been attributed to the DH-(OH-) stretches of waters that are still participating in the hydrogen bonding network of water but have been perturbed by the presence of the interface.
The negative sign implies that their net transition dipole is oriented towards the bulk phase.
While SFG spectra do not show the relative location of these waters at the interface, these can be analyzed by comparison with molecular dynamics (MD) simulations.
In agreement with experiment, the simulations reveal that many interfacial waters have a free OH, while waters further in the bulk (by $\approx$ 1-4 \AA) tend to point towards the bulk phase.\cite{pezzotti20172d, tang2020molecular}.
Lastly, we note that a positive feature in the SFG at lower wavelengths (below 2350 cm$^{-1}$ in Figure \ref{fig:HDSFG}a, below 3200 cm$^{-1}$ in H$_2$O) was later shown to be an experimental artifact. \cite{yamaguchi2015development, nihonyanagi2015accurate} 

Surprisingly, the SFG spectra obtained for planar water-oil interfaces are quite different than that of the air-water interface, as shown for a water-hexane interface in Figure \ref{fig:HDSFG}b\cite{strazdaite2015water}. There is generally a free OH feature, although it is redshifted by $\approx 20 - 40$ cm$^{-1}$ depending on experiment. 
This shift has been attributed to interactions between the oil and the water\cite{scatena2001water, richmond2001structure}, in part because the red shift resembles that for waters in the vicinity of alcohols like butanol\cite{perera2009observation}.
Strikingly, the sign of the hydrogen-bonded OH stretch is now positive, indicating that the net dipole moment in the hydrogen-bound region points away from the bulk water phase, which sharply contrasts with the water-air interface.
Similar results were found for other hydrophobic materials\cite{tian2009structure, de2013analysis, sanders2019heterodyne, yang2020stabilization}, indicating that the change in sign may be general, although the shape of the positive region varies somewhat. The hydrogen-bonded region is also more intense, which Strazdaite et al.\cite{strazdaite2014enhanced} interpreted to mean the hydrogen bonding network is more ordered than the air-water interface. We return to this interpretation later using other surface-sensitive spectroscopic techniques.

We note that the planar water-oil interface is more challenging to prepare than the planar water-air interface, and thus may be more susceptible to artifacts\cite{pullanchery2021charge}. Indeed, ostensibly similar systems have yielded different results\cite{du1994surface, gragson1997comparisons, moore2008integration}, although most recent spectra seem to agree on the shift in the free OH peak and the net dipole of hydrogen-bonded water \cite{strazdaite2015water, tian2009structure, 
de2013analysis,
sanders2019heterodyne, yang2020stabilization}.
However, the SFG spectra disagree with MD simulations, which show that the net dipole of waters points towards the oil phase\cite{xiao2010molecular}, like with the air-water case. This discrepancy may result from ions adsorbed at the interface\cite{tian2009structure}, as discussed below.

\subsection{Interfaces of Oil-Water Emulsions}
\begin{figure}[H]
    \centering
\includegraphics[width=0.995\textwidth]{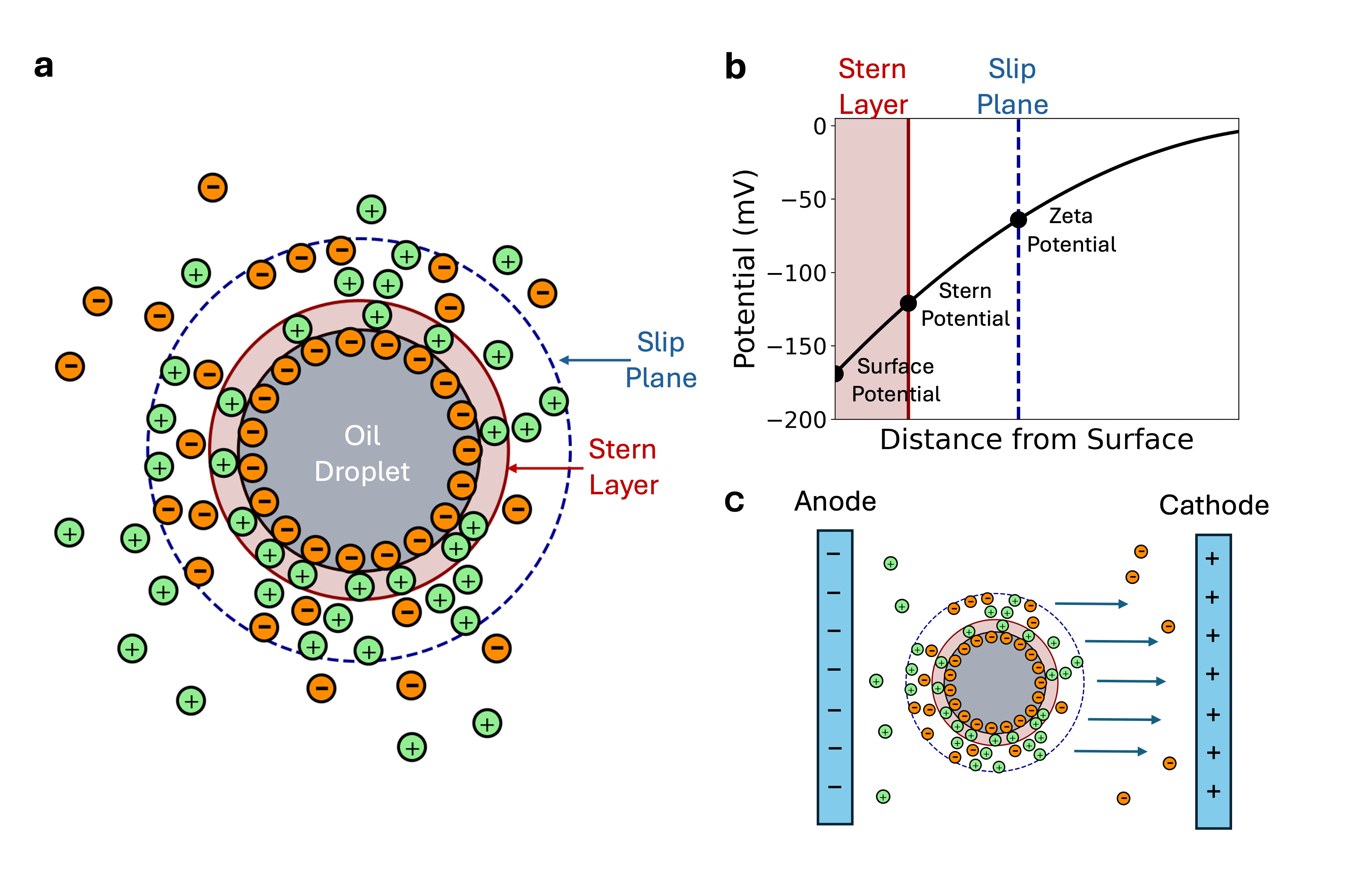}
    \caption{\textit{Molecular view of electrophoresis experiments and the nature of the zeta potential, $\zeta$}. (a) One type of ion is strongly bound to the surface and is surrounded by a cloud of counterions, forming a double layer. The Stern layer defines the immobilized double layer of charge; the slip plane is located further away from the interface such that during electrophoresis due to an applied electric field, ions between the slip plane and the droplet travel with the droplet. (b) An illustration of the potential due to the double layer. The $\zeta$ is the value at the slip plane and is distinct from the Stern and surface potentials. (c) An illustration of electrophoresis; the net charge within the slip plane determines its movement.}
\label{fig:surface_adsorption}
\end{figure}

Like other air-water microdroplets, oil-water emulsions (and gas bubbles in water) display accelerated reactivity.\cite{PhysRevLett.112.028301, vogel2020corona} Oil-water emulsions are examples of electrostatically stabilized droplets, in which the oil (or gas bubbles) carry a net charge that causes the droplets to repel each other, which keeps the emulsion stable by prohibiting their agglomeration into a bulk phase. Figure \ref{fig:surface_adsorption}a shows one possible arrangement of charges in and around an oil droplet in water, but the actual identity of charged species is not fully understood in this case, a topic to which we return to below. Figure \ref{fig:surface_adsorption}b shows a generic organization of charges and counterions that gives rise to a series of electric potentials at varying distances from the hydrophobic surface that helps explain emulsion stability. The surface potential corresponds to the region within which only one type of ion charge is strongly bound to the surface. The Stern layer is defined by a region that includes a cloud of ions of the opposite charge and hence measures the potential due to a charge double layer. The theory of electrophoresis\cite{anderson1989colloid} states that outside the Stern layer are additional ions that also remain immobilized within the ``slip plane'' where the electric potential defines the so-called zeta potential, which can be measured via its relationship to droplet mobility when an electric field is applied. For the case of an electric field applied to gas bubbles or oil droplets in water, the droplets migrate towards the positive electrode indicating that they carry a negative charge (Figure \ref{fig:surface_adsorption}c). The close similarity between the zeta potentials of gas bubbles and oil droplets also indicates that their interfaces are quite similar\cite{agmon2016protons}.

Spectroscopic studies of water-oil emulsions have attempted to provide a molecular understanding of the zeta potential and origin of negative charge. With their much greater interfacial area relative to planar interfaces, emulsions have the added benefit of being less susceptible to artifacts arising from surface contaminants.\cite{shi2024water, pullanchery2020stability,carpenter2019formation} While interfacial spectroscopic measurements of oil-water emulsions have been dominated by SFG, and much has been learned from these experiments, they have a primary disadvantage that the infrared radiation is inevitably attenuated when going through the aqueous medium. This affects the spectrum in a frequency-dependent manner\cite{kulik2020vibrational,carpenter2021effects}, making it difficult to isolate the features just due to the interface. In 2024, Shi et al.\cite{shi2024water}, utilized the Raman multivariate curve resolution technique (Raman-MCR), originally developed for small solutes\cite{perera2008solute}, to obtain the solute-correlated Raman spectrum (SCRS) of oil-in-water emulsions as an alternative to sum frequency experiments. Its primary benefits are its ease of interpretation and ability to compare spectroscopic signatures to the bulk water phase.

\begin{figure}[H]
    \centering
    \includegraphics[width=0.99\textwidth]{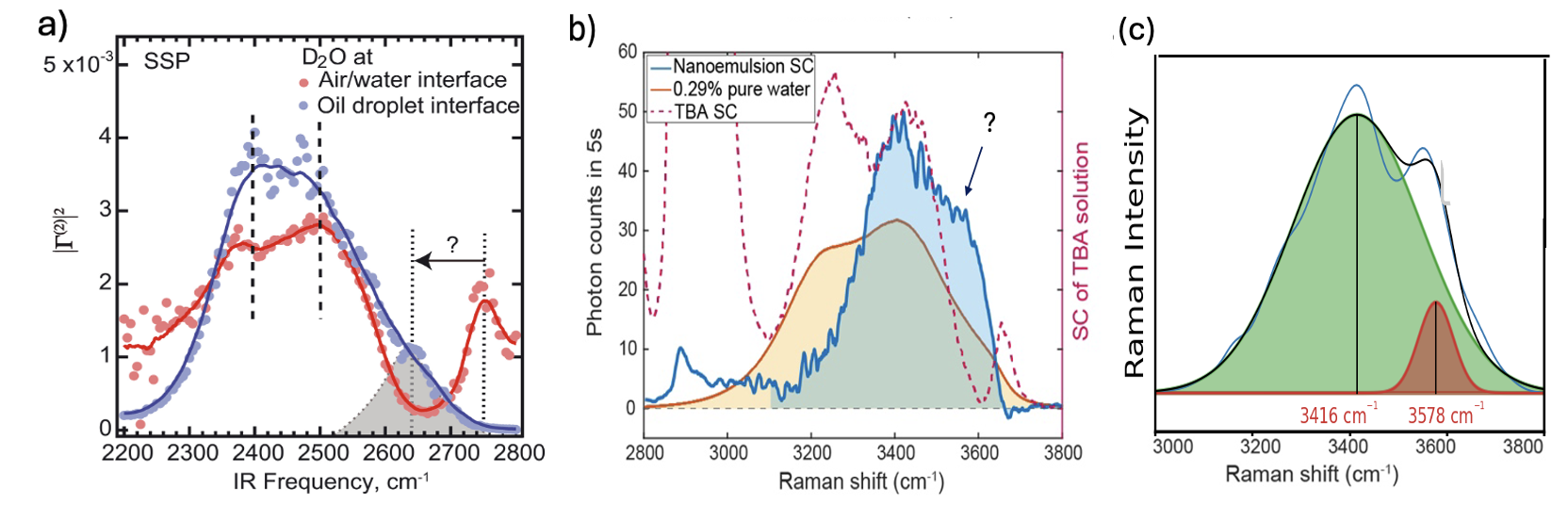}
    \caption{\textit{Experimental and simulated spectra of hexadecane in water emulsions}. (a), the vibrational sum frequency scattering frequency spectra for the oil-water emulsion in blue and the spectra of an air-water interface in red. From Reference \cite{pullanchery2021charge} and reprinted with permission from AAAS. (b) the solute-correlated Raman spectra of the emulsion (blue shading). For comparison, the Raman of pure water (yellow shading) and \textit{tert}-butyl alcohol (dashed red line) are shown. (c) The interfacial spectrum for the AMOEBA model after applying an extra electric field of 92.5 MV/cm to every free OH. Adapted with permission from reference\cite{shi2024water}.}
    \label{fig:emulsion_spectra}
\end{figure}

Figure \ref{fig:emulsion_spectra}(a,b) compares the vibrational sum frequency scattering (VSFS) of hexadecane droplets in waters by Pullanchery et al.\cite{pullanchery2021charge} along with the comparison with the SCRS spectrum of the same oil-water emulsion\cite{shi2024water}. The VSFS and SCRS spectra are seen to resemble each other in the higher frequency range. Specifically, the free OH peak at 2750 (3700) cm$^{-1}$ observed for the air-water interface, or for small hydrophobic solutes, is absent in either spectrum of the emulsion interface. However, in both cases, there is a new shoulder that is a free OH-peak that has been red-shifted due to the interaction with the oil. Interestingly, the lower frequency parts of the VSFS and SCRS spectra differ significantly. In the VSFS spectra, the intensity in this frequency region has increased relative to the water-air interface. In the SCRS spectrum, the opposite has occurred, in that the prominent shoulder at lower frequencies in the spectrum of bulk water has vanished. Considering that the lower frequency shoulder is typically associated with the degree of ordering in the hydrogen bonding network\cite{walrafen1986temperature, sun2013local, pattenaude2018temperature} (although its specific origin is Fermi Resonance\cite{hunter2018disentangling, lacour2023predicting}), the VSFS and SCRS spectra yield opposite conclusions as to degree of order at the interface. This discrepancy may in part be related to VSFS's greater sensitivity to water orientation and Raman-MCR's greater sensitivity to hydrogen bonding strength.

The relative simplicity of Raman-MCR also confers the key advantage that one can compute an equivalent spectrum in theoretical simulation. Figure \ref{fig:emulsion_spectra}c shows the simulated SCRS using the polarizable AMOEBA model in which the lower frequency shoulder's disappearance in the corresponding experimental spectrum is because of weaker hydrogen bonding at the interface, which shifts the Fermi peak off-resonance.\cite{shi2024water} We also find that the simulated spectra of the red-shifted shoulder at higher frequencies can be reproduced by incorporating an additional electric field of 60-90 MV/cm (depending on water model), which is consistent with a zeta potential of $\sim$40-60 mV. 

On a molecular level, the reason why air\cite{quincke1861ueber, mctaggart1914xxxiii, graciaa1995zeta} and oil droplets\cite{carruthers1938electrophoresis, marinova1996charging} dispersed in water have a negative charge and thus migrate towards a positive electrode in an electric field is not fully understood. One possibility is that there are accumulated hydroxide ions bound at the water-oil interface\cite{beattie2009surface}, with a layer of compensating hydronium ions, as the origin of the zeta potential (and depicted in Figure \ref{fig:surface_adsorption}a).\cite{PhysRevLett.112.028301} In contrast, Pullanchery et al.\cite{pullanchery2021charge}, and others previously,\cite{vacha2011orientation, vacha2012charge, poli2020charge, pullanchery2020stability} have proposed that the zeta potential of water emulsions results from charge transfer. In this scenario, many waters on the surface of air-in-water emulsions each transfer a small fraction of their electron density at the edge of the air droplets; in oil-in-water emulsions, the charge is transferred from water molecules to oil molecules. Indeed, an accumulation of charge is observable in MD simulations.\cite{vacha2011orientation, vacha2012charge, poli2020charge} 

However, we find the latter explanation unsatisfactory because it implies an unphysical slip plane. As mentioned above, charges (usually assumed to be ions) on the inner side of the slip plane move with the droplets, while charges on the other side do not. Although an individual water may share a small fraction of its electron density with the droplet, that fraction of electron density is still bound to the water and not the droplet, and thus moves with the water instead of the droplet. The alternative, that small fractions of electron density are bound to the droplet, is distinctly nonphysical because it requires the slip plane to be within individual electrons. And yet the former hypothesis suffers from the disagreement in the experimental literature as to surface acidity or basicity, with more recent studies favoring hydronium, not hydroxide, to be more strongly absorbed at the interface. Finally, some dismiss the idea that there are any inherent molecular interactions involving water and oil, but instead are simply the result of surface impurities.\cite{} This last argument seems the least satisfying for explaining why enhanced chemical reactivity is observed in oil-water emulsions, given that surface contaminants are greatly reduced with this type of droplet preparation. Additionally, any possible impurity cannot originate in the oil itself since emulsions of air bubbles also accumulate a negative charge.

\subsection{Interfacial Adsorption of Organic and Ionic Species}
In microdroplets, reactions rates and thermodynamics will be strongly influenced by their concentration at the interface.
Many reactions reported to be accelerated in microdroplets involve organic reagents and products. Famously, many organic molecules are poorly soluble in water, but they generally adsorb quite strongly to water/hydrophobic interfaces\cite{roth2002adsorption, donaldson2006influence, lemay2023molecular, xiong2020strong2}.
For example, a simulation study\cite{vacha2006adsorption} found that anthracene is roughly 600 times more concentrated at the interface than in bulk water. 
The preference of organic molecules for the water-hydrophobic interface can be understood as a manifestation of the hydrophobic effect. While organic molecules interact favorably with water molecules, their presence in the bulk phase interferes with the highly favorable interactions between water molecules themselves, and thus they segregate toward the surface. Of course, the degree of surface adsorption varies strongly from molecule to molecule\cite{beattie2004pristine}, with soluble molecules having less impetus for the interface.\cite{davies2012interfacial} But the Gibb's adsorption isotherm indicates even molecules that are quite soluble in water, like small alcohols and acids\cite{donaldson1999adsorption} also partition to the interface. This behavior is also seen in MD simulations\cite{ hub2012organic, van2006lifting}. 
As the alkane chain of simple alcohols grows longer, their interfacial propensity also increases substantially\cite{vazquez1995surface}, which can give rise to complex interfaces such as reverse micelles or lipid bilayers.

The interfacial behavior of ions has also been investigated. Like organic molecules, the degree of surface adsorption is correlated with their solvation energy.\cite{hey1981surface, weissenborn1996surface, bastos2016ions} Early theoretical considerations indicated that they should be repelled from the water-air interface\cite{wagner1924oberflachenspannung, onsager1934surface}.
Specifically, charged particles close to the interface between a high-dielectric material (like water) and a low-dielectric material (like a gas or oil phase) are expected to experience an ``image charge repulsion'' from the interface, where the ion experiences a force as if a like-charged ion were on the opposite side of the interface. In simpler terms, this can be understood by considering the fact that ions have strong, long-ranged interactions with water that are interrupted at the interface. However, both experimental and simulation efforts,\cite{jungwirth2002ions, petersen2006nature, jungwirth2008ions, tobias2013simulation, agmon2016protons, bastos2016ions, litman2024surface} indicate that the situation is more complicated. As before, the surface excess of ions can be estimated with the Gibb's adsorption isotherm\cite{weissenborn1996surface}, which shows that many salts are indeed repelled from the surface.
However, some large singly-charged anions are seen to preferentially adsorb to the interface. But the degree of surface adsorption for such ions is not excessive; for example, low-weight alcohols like methanol, which are not known for their surface propensity, adsorb more strongly than these simple ions according to the Gibb's adsorption isotherm\cite{vazquez1995surface}.
We note that the Gibb's adsorption isotherm only reports on the total excess of ions at the interface.
In some case, ions may  adsorb strongly at particular interfacial depth but be depleted overall at the interface, leading to a negative surface excess. \cite{jungwirth2001molecular}

Of particular interest are the roles that hydronium, hydroxide, and electrons play in microdroplets, as well as their relative preferences for the hydrophobic-water interface or bulk phase.\cite{Saykally2013} The influence of the liquid-air interface on the concentration of H$_3$O$^+$ and OH$^-$ ions remains controversial\cite{agmon2016protons, ruiz2020molecular}, with different experiments reaching opposite conclusions. 
One body of evidence comes from oil-in-water emulsions. Their electrophoretic mobility indicates that they carry a negative charge, for pH of 2-4\cite{beattie2009surface, agmon2016protons}. Considering the pH dependence and OH$^-$ being the only negative ion in the solution, this indicates that the surface is rich in OH$^-$. Colussi et al.\cite{colussi2021hydronium} found that trimethylamine only became protonated at pH < 4 when exposed to the microdroplet surfaces in the gas phase, despite becoming protonated at pH > 4 when dissolved in the microdroplets, indicating that H$_3$O$^+$ only becomes present on the surface at lower pH. 
However, at lower pH, surface H$_3$O$^+$ appears to act like a superacid, protonating even very weak bases\cite{colussi2021hydronium}. The movement of oil droplets in water towards a positive electrode has historically been attributed to hydroxide adsorbing to the oily interface.\cite{beattie2009surface}
This mechanism is consistent with the pH-dependence of the zeta potential\cite{carruthers1938electrophoresis, pullanchery2020stability}, whose magnitude is reduced to zero around a pH of 2-4, and the fact that solution pH drops when forming emulsion droplets\cite{beattie2004pristine}. Other interfacial behavior also points towards a negatively charged interface.\cite{beattie2009surface,Mishra2012} Several theoretical explanations exist to justify hydroxide's propensity for the interface, such as its amphiphilic nature\cite{kudin2008water} or its reduction of water's dielectric constant, which decreases dipolar fluctuations\cite{gray2009explanation}. 

In contrast, the Gibb's adsorption isotherm, which is a general method for computing surface excess from surface tension data, indicates that hydroxide is lightly repelled from water-air interfaces.\cite{weissenborn1996surface}
Additionally, SFG spectra suggest that hydronium is more prevalent at the interface, but with the caveat that SFG is more sensitive to acids than bases.\cite{raduge1997surface, baldelli1997sum, tarbuck2006spectroscopic,levering2007observation,  tian2008interfacial, das2019nature, sengupta2018neat, das2019surface}. Furthermore, depending upon the model employed, many MD simulations find that hydronium is at the outermost interface of air-water systems,\cite{Buch2007, hao2022can} whereas hydroxide has no preferential adsorption to interfaces\cite{mucha2005unified, wick2009investigating, tse2015propensity, hub2014thermodynamics, hao2022can,de2023acidity}, only adsorbs weakly\cite{mundy2009hydroxide, yang2020stabilization, bai2016surface}, or stays just below the hydronium interface to form a double layer\cite{litman2024surface} or possibly a triple layer\cite{beattie2009surface} that flips the sign of charge again. Other spectroscopic techniques for probing surface concentrations frequently find the surface to be richer in H$_3$O$^+$ than in bulk water but not richer in OH$^-$.\cite{das2019surface,li2020coordinate} Together this has been interpreted to mean that H$_3$O$^+$ is more surface-active. However, preferential hydronium adsorption at the interface of oil-water emulsions is inconsistent with the zeta potential measurements as patiently pointed out by Beattie and co-workers\cite{beattie2009surface}. To explain these discrepancies, Agmon et al.\cite{agmon2016protons} cite the possibility of differences in the probing depth of spectroscopic techniques, interpretation of experimental quantities like electrokinetic mobility, and the influence of counterions on surface adsorption. Therefore, the molecular origin of the zeta potential remains an unresolved, open question.

Solvated electrons, possibly formed during the preparation of microdroplets, can also significantly influence reactivity. Whether the aqueous electron resides in the bulk or at the air-water interface has been the subject of wide theoretical\cite{herbert2006first,sommerfeld2008model,jacobson2010one,Larsen2010} and experimental\cite{boag1963absorption,Hammer2004,Verlet2005,siefermann2010binding,luckhaus2017genuine,nishitani2019binding,Elg2021,Sopena2024,Jordan2024} interest. In particular, much effort has gone into determining the vertical binding energy (VBE) of the aqueous electron in water clusters and in bulk using ultrafast UV photoelectron spectroscopy. The VBE of the bulk aqueous electron, \ce{e^-_{aq}(b)}, was originally reported to be around 3.3 eV\cite{siefermann2010binding}, but has since been corrected to $3.77\pm0.1$ eV after accounting for kinetic energy loss due to inelastic scattering\cite{luckhaus2017genuine,nishitani2019binding}. In early experimental work, the extrapolation from very small water clusters\cite{Hammer2004} to larger clusters\cite{Verlet2005} suggested that electrons can also bind weakly at the aqueous surface for short times.\cite{siefermann2010binding,Young2012} But like the debate on the surface propensity of the molecular hydronium and hydroxide ions, subsequent SFG and charge transfer to solvent (CTTS) experiments are at odds on whether the electron is partially (electron density exposed to the vapor phase) or fully solvated (density stabilized in a cavity) for an extended air-water interface.\cite{Matsuzaki2016,Nowakowski2016} As summarized by Herbert, the theoretical consensus is that most of the electron density is in the aqueous phase.\cite{Herbert2017} In a recent salvo using electronic absorption spectroscopy, Jordan and co-workers determined that the liberated electron from the surface active phenoxide anion rapidly diffuses into the bulk leaving behind the phenoxide radical at the surface.\cite{Jordan2024} Overall the observation that the electron is fully solvated is consistent with the known fact that \ce{e^-_{aq}} is the only aqueous anion with a positive entropy of hydration\cite{han1990reevaluation}, of $\Delta S_{hyd.}^\circ$ = 118$\pm$20  J/mol/ K$^\circ$. It is also consistent with our recent work showing that hydroxide ions of a microdroplet have a lower vertical ionization energy (VIE) anywhere in the droplet in the presence of excess charge, a process thermodynamically favored by the fully solvated electron.\cite{heindel2022spontaneous}

\section{The Mechanisms of Microdroplet Reactivity}
\subsection{Overview of Mechanisms}

\begin{figure}[H]
    \centering    
    \includegraphics[width=0.99\textwidth]{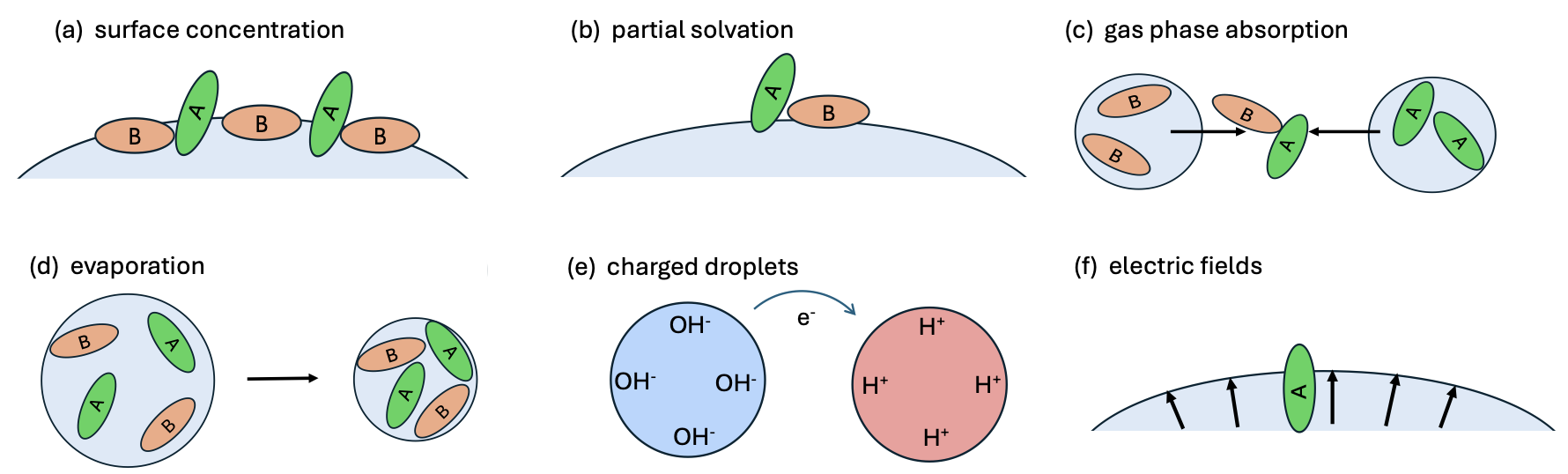}
    \caption{\textit{Possible mechanisms for enhanced droplet reactivity.} For uncharged droplets the (a) thermodynamics of enhanced interfacial concentration, (b) partial solvation that favors transition state stabilization (or reactant destabilization), (c) gas phase absorption, and (d) evaporation are all relevant to explain reactivity in microdroplets. Additionally (e) the creation of a net charge through processes such as contact electrification will layer on top of processes (a)-(d). (f) Interfacial electric fields have received a lot of scrutiny but can be understood as a unifying feature across (a)-(e).}
    \label{fig:mechanisms}
\end{figure}

Having discussed different methods of droplet preparation and the properties of hydrophobic interfaces, we next discuss how those features contribute to reactivity in microdroplets. A number of rate-accelerating mechanisms have been proposed, and here we discuss the most prominent. 
We  illustrate them in Figure \ref{fig:mechanisms}.
We first consider mechanisms that assume the mostly idealized situation in which a droplet is uncharged.

\textbf{Enhanced Interfacial Concentration.} Increased interfacial concentration is perhaps the most straightforward contribution and hence explanation to accelerated reactivity in microdroplets.\cite{ruiz2022disentangling, xiong2020strong2} As described in Section 3, most organic species strongly adsorb to hydrophobic-water interfaces, resulting in a locally enhanced concentration. The higher concentration results in more reagent collisions and accelerated reactivity as per collision rate theory. Importantly, unimolecular reactions will not be accelerated by this mechanism, and indeed, bimolecular reactions appear much more likely to be accelerated in microdroplets.\cite{wei2020accelerated,qiu2021reaction}

Increased interfacial concentrations can alter equilibrium reactant and product concentrations, due to their well-known dependence on the available volume from statistical thermodynamics. Specifically, if reactant species outnumber product species (or vice-versa), the equilibrium product (reactant) concentration will increase in smaller volumes. The imine synthesis reaction examined by Fallah-Araghi et al.\cite{PhysRevLett.112.028301, wilson2020kinetic} provides a lucid example illustrating this model. They examined the bimolecular reaction of an amine and an aldehyde to yield a fluorescent imine in a water-in-oil emulsion. All species had substantial affinities for the interface. Consequently, decreasing microdroplet size increased the equilibrium concentration of the product.

\textbf{Partial Solvation.} The concept of partial solvation\cite{yan2016organic, qiu2021reaction} relies on the observation that many reactions are much faster in the gas phase than in the aqueous phase.\cite{olmstead1977gas, chandrasekhar1984sn2} Thus, if the solvation state at the interface is intermediate between the unsolvated gas phase and the fully solvated aqueous phase, the reaction rates will likely lie between the gas phase and aqueous phase values. Discerning stabilization of the transition state in the gas vs. liquid phase\cite{olmstead1977gas}, is more challenging. Qiu et al.\cite{qiu2021reaction} attributed their observation that only bimolecular reactions are accelerated in microdroplets because these reactions have charge-disperse transition states, which are more stabilized at the interface than in the bulk liquid resulting in a lower barrier. Of course, the acceleration of bimolecular reactions is also expected from an increased interfacial concentration, so it is challenging to distinguish the importance of partial solvation from adsorption thermodynamics. Because it is an atomic-scale mechanism, simulations can help distinguish between different possibilities. To this end, Narendra et al.\cite{narendra2020quantum} used ab initio MD simulations to investigate hydrazone formation from phenylhydrazine and indoline-2,3-dione, finding a reaction path at the interface that had a significantly lower barrier than in the bulk, indicating that partial solvation likely plays a large role in that reaction.

\textbf{Gas-Phase Absorption.} Other explanations point to factors beyond the surface properties of the microdroplets. One possibility is that some chemical reactivity is occurring in the gas phase between reactants that have desorbed from the microdroplet surface. Jacobs et. el.\cite{jacobs2018studying} investigated the production of sugar phosphates, which is reported to be accelerated in microdroplets, finding that the reaction occurred even when the reactants were in separate solutions, and thus concluded that gas-phase reactivity cannot be neglected. Similar conclusions were drawn by Gallo et al.\cite{gallo2019chemical} for a different microdroplet accelerated reaction. 

Recent work has also found that surface enhancement and gas-phase channels are kinetically coupled to other reactions occurring in a droplet and that the kinetics of such processes accelerate sigmoidally as droplet size decreases, indicating the importance of the surface.\cite{li2023size,li2024enhanced} It has also been observed that when organic acids are present, formation and desorption of \ce{HNO_3} and \ce{HCl} from the interface contribute to the droplet chemistry by pH modulation.\cite{angle2023direct} In fact, droplet pH strongly modulates uptake of many gaseous species which can accelerate interfacial reactions. One example is the interplay of \ce{O_3} uptake, pH, and iodide oxidation occurring at either the surface or in the bulk of a levitated droplet\cite{prophet2024distinguishing}. Taken together, these studies indicate that gas-phase reactivity can be the dominant contribution to accelerated reactivity in some cases and, through kinetic coupling with gas-phase channels, can accelerate reactions which might otherwise occur slowly.

\textbf{Droplet Evaporation.} This raises the possibility that, in general, droplets are evaporating over the course of the experiment, thus concentrating the reactants and accelerating the reaction. While evaporation can certainly occur, it is difficult to ascertain its relevance. In one set of experiments, Lai et al.\cite{lai2018microdroplets} found that changing the distance that the droplets must travel to the detector had a limited effect on the reaction profile, indicating that limited evaporative concentration of reactants was occurring. On the other hand, when studying a reaction in which the reagents had limited surface propensity, Chen et al.\cite{chen2023role} found that solvent evaporation dramatically influenced reaction rate. 

\subsection{Influence of Microdroplet Charge on Enhanced Reactivity.} Many redox reactions have been reported to be greatly accelerated in microdroplets, as summarized in the reviews of Jin et al.\cite{jin2023spontaneous} and Vannoy et. al.\cite{vannoy2024electrochemical} These include the reduction of the pyridyl anion\cite{zhao2022sprayed}, which is thought to very unstable under normal conditions\cite{nenner1975temporary}, and the reduction of various metals \cite{he2022vapor, yuan2023spontaneous}, yielding the formation of various complexes and even nanomaterials\cite{lee2018spontaneous}. Many of these reactions are thermodynamically unfavorable in neutral microdroplets. 
A prominent example is the production of hydrogen peroxide from water:
\begin{equation}
    \ce{H$_2$O(\ell) <=> 1/2H_2O_2(aq) + 1/2H_2(g)},
    \label{eq:H2O2}
\end{equation}
which has equilibrium constant $ < e^{-40}$. Nonetheless, a number of research labs\cite{musskopf2021air, mehrgardi2022sprayed,nguyen2022revisiting} report its formation in aqueous microdroplets, generating much controversy regarding the underlying amount of H$_2$O$_2$ and the mechanism by which it happens.\cite{lee2020condensing, musskopf2021air, chen2022water, gallo2022formation,eatoo2024busting} 

\begin{figure}[H]
    \centering
    \includegraphics[width=1.0\textwidth]{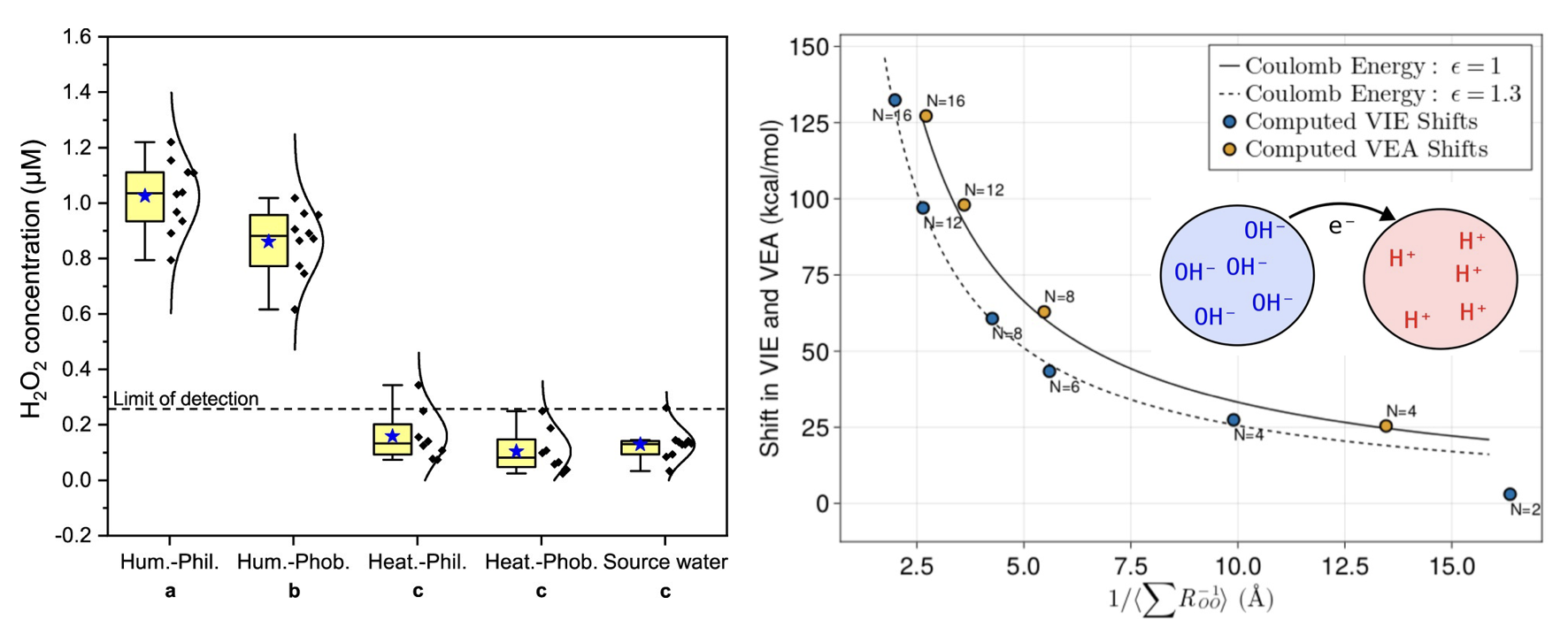}
    \caption{\textit{The formation of H$_2$O$_2$ in droplets driven by destabilization of OH$^-$ and H$_3$O$^+$ by promoting electron transfer with excess charge.} (a) hydrogen peroxide generated from condensing water vapor using gentle heating (Heat) compared to using an ultrasonic humidifier (Hum) and then condensed on hydrophobic (Phob) or hydrophilic (Phil) surfaces. Reproduced with permission  from\cite{musskopf2021air}. (b) the blue points give the VIE energies for OH$^-$; the yellow points give the VEA energies for H$_3$O$^+$.
    In the insert we show a proposed mechanism for how charged droplets are formed and for how redox chemistry occurs. Larger volumes of water separate into smaller microdroplets, for which the amount of OH$^-$ and H$_3$O$^+$ is not evenly divided. Electrons are then destabilized within the negative droplets but stabilized within the positive droplets. Adapted with permission from reference\cite{heindel2024role}.}
    \label{fig:h2o2_gen}
\end{figure}

Hydrogen peroxide's formation might be justified if it exhibited a large binding affinity to the interface. However, its binding affinity to water-air interfaces is only around -1 kcal/mol\cite{vacha2004adsorption, martins2017reaching, donaldson2022}, so its formation clearly depends on factors beyond the thermodynamics of air-water interfacial absorption. This point is further demonstrated by the fact that the amount of H$_2$O$_2$ formed in microdroplets varies depending upon how the microdroplets are made. As shown in Figure \ref{fig:h2o2_gen}a, Musskopf et. al.\cite{musskopf2021air} found that the H$_2$O$_2$ concentration in microdroplets formed from condensing water vapor depended upon whether the vapor was generated from an ultrasonic humidifier or by gentle heating. Only the droplets generated from the ultrasonic humidifier showed H$_2$O$_2$ above the detection limits of their analyzer (Figure \ref{fig:h2o2_gen}a). Another study found substantially higher concentrations of H$_2$O$_2$ using an ultrasonic mist maker as well\cite{nguyen2022revisiting}. 

How might the observed chemical reactivity of not only H$_2$O$_2$ but many redox reactions be explained? One immediately relevant and underappreciated property of microdroplets for redox chemistry is their charge. Experimentally, both negatively charged and positively charged droplets are present in water sprays\cite{maze2006negative, zilch2008charge, lin2023size}. The ESI experiment itself, for which the largest rate accelerations have been reported\cite{banerjee2017can, wei2020accelerated}, relies directly on the production of highly charged droplets\cite{banerjee2012electrospray}. Furthermore, droplets can acquire charge in other situations, and it even appears that producing uncharged water microdroplets is the exception. For example, water streams are known to spontaneously acquire charge\cite{faraday1843iii}.
Various levitated droplets\cite{abdelaziz2013green, wang2021study} spontaneously acquire a charge, and sonication and even pipetting are both capable of producing a charged droplet\cite{choi2013spontaneous, nauruzbayeva2020electrification}. Evidence in favor for the importance of charge is the lack of reactivity in droplets prepared by heating and cooling, as seen in Figure \ref{fig:h2o2_gen}a, which are thought to have minimal charge.
Likewise, Banerjee et al.\cite{banerjee2015syntheses} found that increasing microdroplet charge resulted in greater reactivity. 

Certainly charged droplets substantially alter the thermodynamics of redox reactions relative to neutral droplets. Colussi has made the claim that both hydroxide and hydronium ions are preferentially solvated at the surface of sprayed (i.e. charged) droplets, and that they have a dramatically lower hydration enthalpy that explains the observation of \ce{H2O2} formation.\cite{colussi2023mechanism} This work simply assumes surface activity of \ce{H_3O^+} and \ce{OH^-} and estimates the enthalpy of this surface state from gas phase thermodynamic tables.\cite{colussi2023mechanism} However, the proposed mechanism is invalidated by our recent work based on quantitative simulations and theory using a thermodynamic cycle for nanodroplets with net charge.\cite{heindel2024role} We find that the ion spatial distribution in net charged environments is weakly perturbed with respect to the ion distribution when a droplet has just one charge, and furthermore is not a surface effect.\cite{heindel2024role} By computing the vertical ionization energies (VIEs) of hydroxide and vertical electron affinities (VEAs) of hydronium in nanodroplets that have an excess amount of hydroxide and hydronium, the VIEs and VEAs shift substantially.\cite{heindel2024role} The magnitude of the shift closely matches an unscreened Coloumb's law between like-charged ions as shown in Figure \ref{fig:h2o2_gen}b. This observation and the computed hydration enthalpies of each ion provide a direct connection to the reaction thermodynamics explaining how solvated electrons and hydroxyl radicals can be produced spontaneously in sprayed droplets. Furthermore, the excess droplet charge that is needed for favorable thermodynamics is well below the Rayleigh limit. Finally, using scaling arguments to reach the micron scale, the thermodynamics only become more favorable for a particular fraction of the Rayleigh limit. It stands to reason that other unfavorable redox reactions may also become favorable under such conditions.

As with gas bubbles and oil emulsions in water, the fundamental origin behind the charging of air-water droplets and oil-water emulsions is unclear. It is likely related to contact electrification\cite{chen2022water, vannoy2023pluses}, which refers to the generic observation that charge can be exchanged between two materials in contact, leaving both with a net charge.\cite{lowell1980contact,xu2018electron,wang2019origin}  While CE has been observed to occur in all combinations of gas, liquid, and solid contacts\cite{wang2019origin}, the most relevant types of CE for microdroplet reactivity are gas-liquid, liquid-liquid, and liquid-solid. Unfortunately, the theory of CE remains incomplete and poorly understood especially when liquids are involved.\cite{wang2019origin, sun2021understanding} Nonetheless, recent work studying sonication of water in contact with fluorinated ethylene propylene (FEP) found that \ce{H2O2} is generated by two pathways.\cite{berbille2023mechanism} The first involves the reduction of \ce{O2} to \ce{O_2^.-} by electron transfer from either FEP or \ce{OH^-}. The second involves oxidation of \ce{H2O} to \ce{H_2O^+} which immediately decays to \ce{OH^.} or direct oxidation of \ce{OH^-} to \ce{OH^.}. This type of process is referred to as contact-electro-catalysis\cite{wang2024contact} because it produces \ce{H2O2} via an oxidative and reductive pathway such that the FEP surface acts as both a source and sink of electrons.

Analogously, charge-stabilized oil-water emulsions can be seen to arise from liquid-liquid CE where sonication plays the role of repeatedly driving the contact between the oil and water. In the aforementioned study, the charges are electrons originating from the solid or liquid\cite{berbille2023mechanism}, while in the case of an oil-water emulsion it is believed that the surface charges are molecular ions, usually identified as \ce{OH^-}.\cite{beattie2009surface} There is also direct evidence of the CE process between a gas and levitated liquid droplet in which the levitated droplet accumulates a positive charge.\cite{wang2021study} This is interpreted to mean that electrons are transferred from the droplet to gas molecules.

Droplet charges can impact reactivity in other ways as well. For clarity, we note that the pH is defined as the logarithm of the activity of H$^+$, so the pH at the interface necessarily equals the value in bulk\cite{colussi2018can}. Nonetheless, the concentration of H$_3$O$^+$ or OH$^-$ may be enriched at the surface such that the water interface has been called both a "superacid"\cite{enami2010superacid, colussi2021hydronium}, and a "superbase"\cite{tabe2014surface, colussi2021hydronium, huang2021accelerated}. Consequently, while uncharged planar air-water interfaces may not behave like superacids or superbases, the interfaces of charged microdroplets might. Enhanced reactivity has also been reported in water emulsions. Specifically, hydrogen peroxide may form spontaneously in water droplets dispersed in another phase\cite{krushinski2024direct}, and the droplets may be generally useful for electrochemical reactions\cite{vogel2020corona}.
Considering that emulsion droplets also carry a charge, their enhanced reactivity may have a common origin to that of aerial microdroplets.
At the same time, oil-water emulsion droplets all have the same charge whereas aerial droplets can have either sign of charge. Thus, reactivity in emulsions may be limited to certain types of redox reactions compared to aerial microdroplets prepared by ESI or gas nebulization. In either case there is enhanced redox reactivity in droplet preparations that create charged droplets relative to neutral droplets.

\subsection{Interfacial Electric Fields}
\begin{figure}[H]
    \centering
    \includegraphics[width=0.85\textwidth]{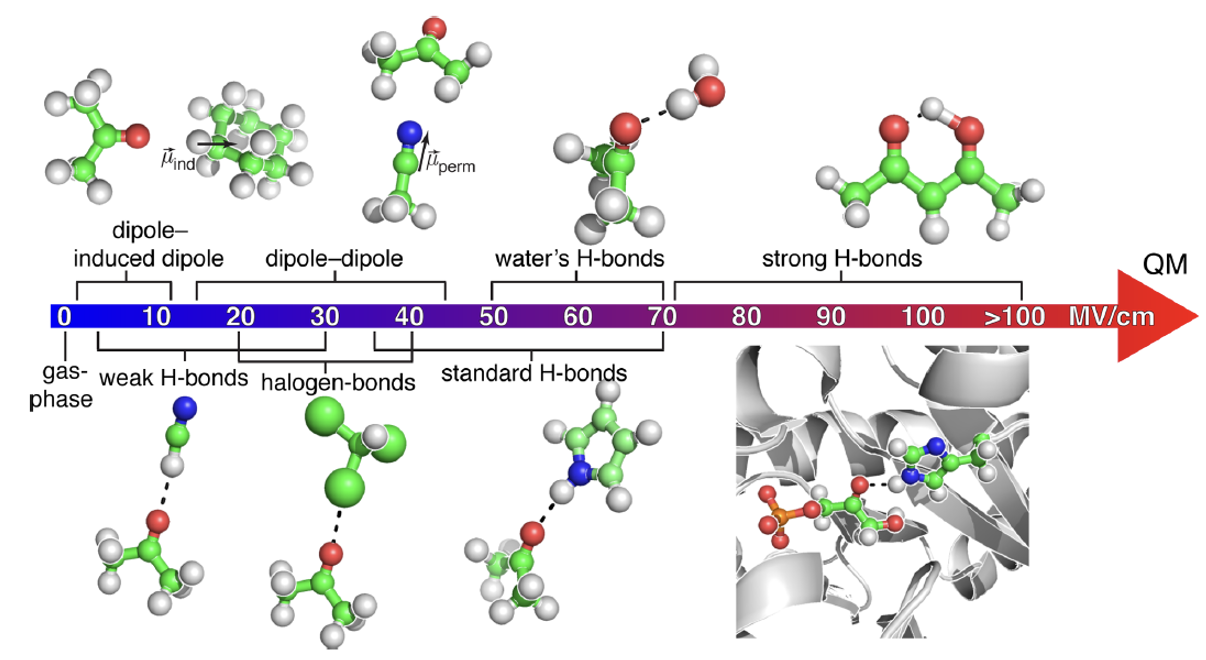}
    \includegraphics[width=0.85\textwidth]{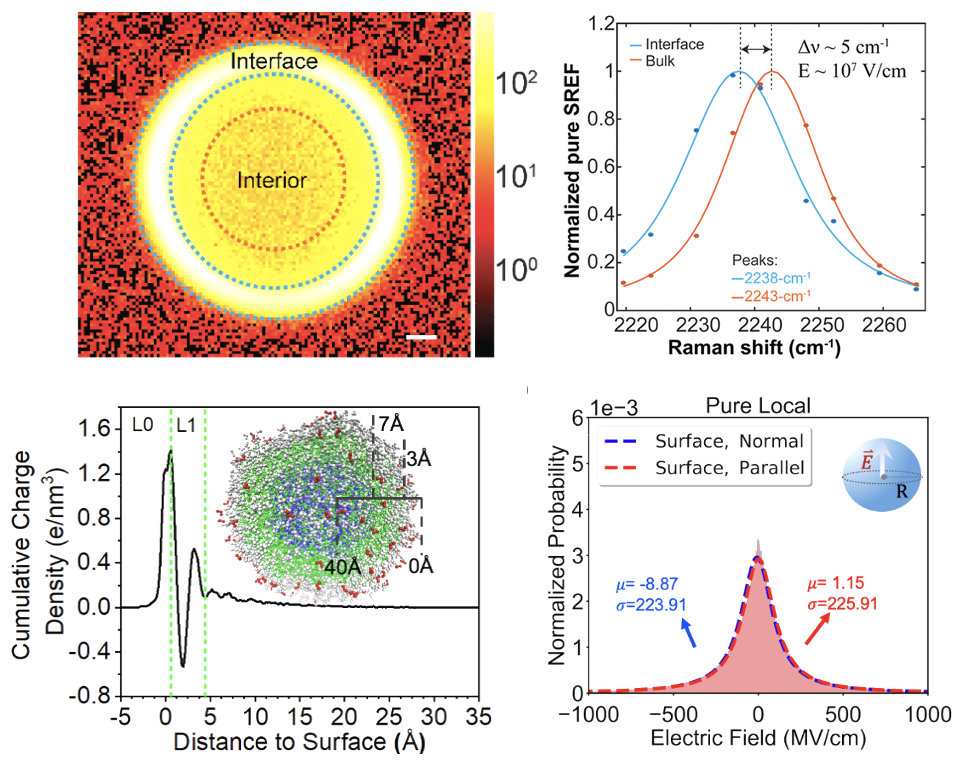}
    \caption{\textit{Electric fields measured at water interfaces.} (a) Stimulated Raman excited fluorescence (SREF) imaging of water microdroplets in oil. (b), the frequency of the nitrile stretch of rhodamine 800 at the interface of a water-in-oil emulsion and in the bulk. (c) cumulative charge density as simulated with a reactive force, ReaxFF/C-GeM\cite{leven2020reactive}, for the air-water interface of nanodroplets showing fluctuations over the nm lengthscale. (d) Lorentzian distribution of electric fields evaluated at the air-water interface using ReaxFF/C-GeM.
    Adapted with permission from references\cite{xiong2020strong1, hao2022can}.}
    \label{fig:electric_fields2}
\end{figure}

Like electrostatic preorganization known to occur in enzymes\cite{Warshel1978,Kamerlin2010, Welborn2018,Li2021}, microdroplets also have an organized water structure at their interfaces.  that depends on how the droplet is prepared (charged or uncharged), what type of water interface is formed (air, oils, or solid surfaces), and what types of surface sensitive adsorbants accumulate at the interface. Thus measured electric fields can be regarded as a unified and quantitative gauge of how these many molecular interactions integrate at formed water interfaces as demonstrated in Figure \ref{fig:electric_fields2}a.\cite{fried2017electric,Welborn2018} There are net field strengths of 10 to 100 MV/cm that arise from intermolecular interactions at the molecular surfaces, as well as much larger fields originating from the large potential gradients inside the electron density of molecules\cite{Yesibolati2020, Kathmann2021}. We are typically interested in the former for understanding chemical reactivity since reactant molecules cannot overlap strongly with the internal electron density of other molecules. Mechanistically, if the electric field is parallel to a reaction coordinate axis, the electric field will promote ionization along the bond\cite{shaik2016oriented} and consequently reduce the energy required for the bond to break. Thus, electric fields have been put forth as an explanation for a wide array of accelerated reactions in enzymes\cite{Welborn2019}, synthetic catalysts\cite{li2020catalytic,Maitra2024}, electrocatalytic surfaces\cite{Welborn2018}, and most recently for microdroplet chemistry.\cite{hao2022can} 

Water, with its large dipole moment, generates large electric fields. While molecules in bulk water can experience large fields of around 100 - 300 MV/cm \cite{smith2005unified}, rotational averaging creates fields that are very short-ranged and short-lived such that projections along reacting bonds are largely very rare events. However at the interfaces formed by water the reactant molecules can experience electric fields that are more persistent both in direction and in time. Hence there is great interest in estimating the surface field strengths experienced by reactant solutes that might drive chemical reactivity. In 2020, Xiong et. al.\cite{xiong2020strong1} used the rhodamine 800 probe to image the electric field at the surface of a water-in-oil emulsion using stimulated Raman excited fluorescence (SREF) (Figure \ref{fig:electric_fields2}b). They found that the frequency of the probe's nitrile group, which is sensitive and calibrated to the local electric field, was shifted at the interface by 5 cm$^{-1}$, which corresponds to a field strength of $\approx$ 10 MV/cm (Figure \ref{fig:electric_fields2}c). Likewise, Hao et al.\cite{hao2022can} examined the cumulative charge density, surface potential, and electric fields for a water nanodroplet, quantities experienced by the OH vibrational stretches at the interface relative to bulk (Figure \ref{fig:electric_fields2}d). They found that the mean electric field at the interface is close to $\approx$ 10 MV/cm, in agreement with the SREF experiment. This average electric field value is likely too small to drive chemical reactions even with perfect alignment along a breaking bond, which requires at least another order of magnitude in field strength. But  the soft interface can fluctuate to create a Lorentzian spread of field values that would certainly be large enough to influence reactivity with reasonably high probability as seen in Figure \ref{fig:electric_fields2}e.\cite{shaik2016oriented, shaik2020electric,hao2022can}. 

The presence of net charge of an air/water droplet or oil-water emulsion can lead to a greater surface potential and electric field, but also manifests as electric fields that are longer-ranged and extend over the entirety of the microdroplet as shown by Chamberlayne and Zare.\cite{chamberlayne2020simple}. Microdroplets in emulsion systems also carry a charge, which can increase the strength of electric fields for these hydrophobic-water interfacial systems as well. In a well-cited study, the surface charge density (which must be sufficient to stabilize the emulsion) is obtained by measuring the size of the emulsion droplets by electroacoustics and the quantity of NaOH required to keep the pH constant during homogenization.\cite{beattie2004pristine} Hence, the measured pH of the solution drops so that the surface charge density of negative ions can be measured by titration.\cite{beattie2004pristine,creux2009strong} For hexadecane, the net surface charge density is approximately -4.6 $\mu$C/cm$^{-2}$ at neutral pH. Application of Gauss' law for a 125 nm droplet using this experimentally derived charge density estimate yields an electric field of $\approx$55 MV/cm at the surface of oil-water emulsions. In the SCRS spectra described in Section 2\cite{shi2024water}, we estimated the electric field at the interface due to the zeta potential in an oil-in-water emulsion to be around 40-90 MV/cm, which is substantially larger than the values reported at uncharged interfaces. We thus suspect that fields that arise due to microdroplet charge play a more important role in accelerating microdroplet chemistry than the fields inherent to uncharged water interfaces.

\section{Conclusions}
Water microdroplets show promise in enabling a diverse array of reactions, many of which are quite unfavorable in a homogeneous phase, and providing a "clean" synthesis platform that is being explored for the ability to extend them to the industrial scale. To achieve this requires that we confront the complexity of aqueous microdroplets, which is underscored by deficiencies in our understanding of water/hydrophobic interfaces in general. In this perspective  we have reviewed this understanding, emphasizing insights from recent work while also highlighting areas of disagreement.

One aspect of microdroplets that we are confident is relevant for their reactivity is their charge. Recent work of ours and others demonstrates that charge can dramatically change the reaction thermodynamics, making unfavorable reactions favorable, although further work is needed to establish the origin of the charge. Therefore, experiments able to explicitly connect the zeta potential and interfacial electric fields to specific molecular species will be tremendously insightful. Additionally, both experimental and theoretical insights into the mechanisms of contact electrification at liquid-gas, liquid-liquid, and liquid-solid interfaces are needed to tell the complete story of microdroplet reactivity. To this end, UV-vis spectroscopy of oil-water emulsions could help elucidate the presence or absence of radical anions produced by sonication since hemi-bonded systems tend to have intense UV-vis absorption bands between 200-400nm. The importance of contact electrification for microdroplet reactivity is that the kinetic energy of molecules is transformed into large charge separations. In this sense, contact electrification can be thought of as the original source of energy which is used to drive the myriad reactions we have discussed above. Therefore, elucidating the mechanisms of contact electrification involving liquids is critical for designing efficient reactions in charged microdroplets.

Throughout this perspective, we have argued that there are fundamental similarities between sprayed microdroplets and oil-water emulsions. This argument has been made on the grounds that each system has regions of net-negative and net-positive charge and these charged regions should alter the thermodynamics of chemical reactions in fundamentally similar ways. For these reasons, we also expect that there are both experimental and theoretical connections to be made between microdroplet chemistry and the chemistry which occurs at electrochemical interfaces.\cite{Welborn2018,chamberlayne2020simple} 

It should also be emphasized that theoretical models such as Gouy-Chapman\cite{gouy1910,Chapman1913} or theory for dielectric saturation\cite{Debye1929,Onsager1936,Kirkwood1939,Booth1951} starts with the assumption of system neutrality that is violated by the preparations that create charged droplets, preparations that exhibit the greatest level of rate accelerations. Reformed theoretical models of this kind would allow the effect of excess charge on thermodynamics and electronic structure to be studied routinely while also providing insights into ion distribution in charged systems. Additionally, hydroxyl radicals are centrally important in many reactions which occur in aqueous microdroplets, but there are few reactive models of \ce{H_3O^+}, \ce{OH^-}, and relevant radical species, that are currently available.\cite{Knight2012,Leven2020} Microdroplets are thus an excellent motivation for the development of new reactive force fields.\cite{Leven2021} 

We have summarized the diverse and fascinating chemistry which occurs in microdroplets and the mechanisms by which these reactions are thought to occur. In neutral systems, we believe that a complex combination of interfacial electric fields, partial solvation, gas-phase absorption, and concentration enhancements are all mechanistically relevant. Many of the most interesting applications of microdroplets involve accelerating redox chemistry. For such reactions, we argue that excess charge is the key variable. This excess charge can be dispersed throughout the system, as with sprayed droplets, or concentrated in a specific region as with the electric double layer stabilizing oil-water emulsions. 

\section*{Acknowledgments}
We thank the Air Force Office of Scientific Research through the Multidisciplinary University Research Initiative (MURI) program under AFOSR Award No. FA9550-21-1-0170 for the microdroplet chemistry application. We also thank the CPIMS program, Office of Basic Energy Sciences, Chemical Sciences Division of the U.S. Department of Energy under Contract DE-AC02-05CH11231 for the Raman monomer-field theory. We thank the National Science Foundation under Grant CHE-2313791 for the reactive force field used in modeling microdroplets.

\section*{Author contributions} J.H., R.A.L. and T.H-G. conceived the theme and wrote the manuscript, and all authors contributed to all insights through extensive discussion.

\section*{Competing interests} All authors declare no competing interests.

\bibliographystyle{unsrtnat}
\bibliography{references}

\end{document}